\def\comment#1{}

\newcommand{\beg}{\begin{eqnarray}}
\newcommand{\eee}{\end{eqnarray}}

\documentstyle[prl,aps,amsfonts,psfig,multicol,epsf]{revtex}
\def\cm#1{}

\begin{document}
\title{ 
Hidden symmetry and knot solitons in a charged two-condensate Bose system}
\author{
Egor Babaev$^1$
\thanks{Email: egor@teorfys.uu.se \  http://www.teorfys.uu.se/people/egor/  } 
Ludvig D. Faddeev$^{2}$
\thanks{Email: faddeev@pdmi.ras.ru} 
and Antti J. Niemi$^1$
\thanks{Email: antti.niemi@teorfys.uu.se \ http://www.teorfys.uu.se/people/antti/}}
\address{
$^1$Institute for Theoretical Physics, Uppsala University 
Box 803, S-75108 Uppsala, Sweden \\
$^2$St.Petersburg Branch of Steklov Mathematical
Institute,
Russian Academy  of Sciences, Fontanka 27 , St.Petersburg, 
Russia 
}
\comment{\\
}

\maketitle
\begin{abstract}
We show that a charged two-condensate Ginzburg-Landau
model or equivalently a Gross-Pitaevskii functional for 
two charged Bose condensates, can be mapped onto
a version of the nonlinear $O(3)$ $\sigma$-model. This 
implies in particular that such a system possesses a hidden 
$O(3)$ symmetry and allows for 
the formation of stable knotted solitons. 
\end{abstract}
\newcommand{\la}{\label}
\newcommand{\aaa}{\frac{2 e}{\hbar c}}
\newcommand{\Pfaff}{{\rm\, Pfaff}}
\newcommand{\kA}{{\tilde A}}
\newcommand{\G}{{\cal G}}
\newcommand{\cP}{{\cal P}}
\newcommand{\M}{{\cal M}}
\newcommand{\E}{{\cal E}}
\newcommand{\btd}{{\bigtriangledown}}
\newcommand{\W}{{\cal W}}
\newcommand{\X}{{\cal X}}
\renewcommand{\O}{{\cal O}}
\renewcommand{\d}{{\rm\, d}}
\newcommand{\bfi}{{\bf i}}
\newcommand{\e}{{\rm\, e}}
\newcommand{\bfx}{{\bf \vec x}}
\newcommand{\bfn}{{\bf \vec n}}
\newcommand{\bfE}{{\bf \vec E}}
\newcommand{\bfB}{{\bf \vec B}}
\newcommand{\bfv}{{\bf \vec v}}
\newcommand{\bfU}{{\bf \vec U}}
\newcommand{\bfp}{{\bf \vec p}}
\newcommand{\f}{\frac}
\newcommand{\bfA}{{\bf \vec A}}
\newcommand{\non}{\nonumber}
\newcommand{\be}{\begin{equation}}
\newcommand{\ee}{\end{equation}}
\newcommand{\ba}{\begin{eqnarray}}
\newcommand{\ea}{\end{eqnarray}}
\newcommand{\bastar}{\begin{eqnarray*}}
\newcommand{\eastar}{\end{eqnarray*}}
\newcommand{\half}{{1 \over 2}}
\begin{multicols}{2}
\narrowtext
For over forty years there has been a wide
interest in condensed matter systems with 
several coexisting Bose 
condensates \cite{mult}. Here we shall
investigate the physically important example
of two charged condensates together with their
electromagnetic interaction. This system is described
by a Ginzburg-Landau model with two flavors of 
Cooper pairs. Alternatively, it relates to
a Gross-Pitaevskii functional with two {\it charged} 
condensates of tightly bound fermion pairs, or some 
other charged bosonic fields.
Such theoretical models have a wide range of applications 
and have been previously considered  in
connection of two-band superconductivity. Indeed, these 
models  describe
superconductivity in transition metals \cite{mult,tg}.
The presence 
of two condensates
 has 
been observed in experiments on  $Nb$, $Ta$ and $V$
as well as in $Nb$-doped $Sr Ti O_3$ \cite{transi}.
More recently the renewed interest  to two-gap superconductivity 
was sparked by discovery of the the
two-band superconductor with surprisingly high 
critical temperature $Mg B_2$ \cite{mgb}.
It has also been argued in \cite{ashc1} that under 
certain conditions liquid metallic hydrogen might
allow for the coexistence of superconductivity 
with both electronic and protonic Cooper pairs.  
In a liquid metallic deuterium a deuteron superfluidity 
may similarly coexist with superconductivity of electronic 
Cooper pairs \cite{ashc2} (see also \cite{ashc3}). 
 
Here we shall be particularly interested in an {\it 
exact} equivalence
 between the two-flavor 
Ginzburg-Landau-Gross-Pitaevskii (GLGP) model
and a version of the nonlinear O(3) $\sigma$-model
introduced in \cite{fadde}. We expose this equivalence
 by 
presenting an exact, explicit change of variables between the
two models. The model in 
\cite{fadde} is particularly interesting since 
it describes topological excitations 
in the form of stable, finite length knotted 
closed vortices \cite{nature}. The equivalence
then implies that a system with two charged 
condensates similarly supports topologically nontrivial,
knotted solitons. Previously it has been argued 
that these topological defects could 
play an important role  in high energy 
 physics \cite{fadde}-\cite{plasma}. 
The purpose of the present Letter is to 
discuss the condensed matter counterparts.  

A system with two electromagnetically coupled, oppositely charged Bose 
condensates can be described by a two-flavour (denoted by $\alpha =1,2$) 
Ginzburg-Landau or Gross-Pitaevskii (GLGP) functional,
\beg
&&
F =\int d^3x \ \biggl[ \frac{1}{2m_1} \left| \left(\hbar \partial_k +
i \f{2e}{c} A_k\right) \Psi_1 \right|^2 + 
\nonumber \\
&& 
\frac{1}{2m_2}  \left| \left(\hbar \partial_k -
i \f{2e}{c} A_k\right) \Psi_2 \right|^2 
+ { V} (|\Psi_{1,2}|^2)
+ \frac{{\bf B}^2}{8\pi},
\biggr]
\la{act}
\eee
where we take
\[
{V} (|\Psi_{1,2}|^2)=-b_\alpha|\Psi_\alpha|^2+ 
\frac{c_\alpha}{2}|\Psi_\alpha|^4
\]
Here we shall consider the general case where the two
condensates are characterized by different effective masses 
$m_\alpha$, different coherence 
lengths $\xi_\alpha=\hbar/\sqrt{2m_\alpha b_\alpha}$ and 
different concentrations
$N_\alpha=<|\Psi_\alpha|^2>={b_\alpha}/{c_\alpha}$ \cite{rema}. 
The properties of the corresponding model with a single charged 
Bose field are well known. In that case the field degrees of freedom
are the massive modulus of the single complex order parameter
and a vector field which gains a mass due to the Meissner-Higgs effect.
The important
property of the present GLGP model is that the two charged 
fields are not independent but there is a nontrival coupling
which is mediated by the electromagnetic field. This implies that
we have a nontrivial, hidden topology which becomes obscured 
when we represent the model in the variables (\ref{act}).
In order to expose the topological structure we introduce
a new set variables, involving a massive field $\rho$ 
which is related to the densities 
of the Cooper pairs and a three-component 
unit vector field $\bfn$. The important 
feature of these new variables is their 
gauge invariance: Neither the relative phase of the condensates 
$\Psi_1 $ and $\Psi_2$ nor the gauge field $\bf A$ enters 
in the free energy functional
when represented in these new variables. 

We start by introducing 
variables $\rho$ and $\chi_{1,2}$ by
\[
\Psi_\alpha = { \sqrt{2m_\alpha}} \ \rho \chi_\alpha
\]
where 
the complex $\chi_\alpha=|\chi_\alpha| 
e^{i \varphi_\alpha}$  are chosen so that 
$|\chi_1|^2+|\chi_2|^2=1$.
The  modulus $\rho$ then has the following expression: 
\[
\rho^2=\f{1}{2}\left( \frac{|\Psi_1|^2}{m_1} + \frac{
|\Psi_2|^2}{m_2} \right)
\]
In terms of the variables $\rho$ and $\chi_\alpha$ 
the free energy density in 
(\ref{act}) reads as follows:
\beg
F &=& \hbar^2(\partial \rho)^2 + \hbar^2\rho^2 
|(\partial_k+  i\kA_k) \chi_1|^2 +
\nonumber  \\
&& 
\hbar^2 \rho^2|(\partial_k- i\kA_k) \chi_2|^2 
+ \frac{{\bf B}^2}{8 \pi} +
{V}
\label{act2}
\eee
where we denote $\kA_k = \aaa A_k$, and    $\partial$ (without an index) is the 
ordinary $\nabla$ operator. The standard gauge invariant 
expression for the supercurrent density 
\beg
{\bf J }&=& 
\frac{i \hbar e}{m_1}
\left\{\Psi_1^*\partial \Psi_1-
\Psi_1 \partial \Psi_1^*\right\}
-\frac{i \hbar e}{m_2}
\left\{\Psi_2^*\partial \Psi_2-
\Psi_2 \partial \Psi_2^*\right\} \nonumber \\
&-&\frac{4e^2}{c} \left( \f{|\Psi_1|^2}{m_1} 
+\f{|\Psi_2|^2}{m_2} \right){\bf A}
\la{A}
\eee
becomes in these new variables 
\beg
{\bf J} = 4 \hbar  e \rho^2 \left[ \f{\bf j}{2} - {\bf \kA}\right].
\label{jjj}
\eee
where 
\[
{\bf j}= i  [ \chi_1\partial \chi_1^* - \chi_1^*\partial \chi_1
- \chi_2\partial \chi_2^* + \chi_2^*\partial \chi_2].
\]

We introduce a gauge-invariant vector field 
$\vec{\sf C}$, directly related to the supercurrent density by
\be
\vec{\sf C}=\frac{{\bf J}}{e \rho^2} 
\ee
We then rearrange the terms in (\ref{act2}) as follows:
We add and subtract from (\ref{act2}) a term 
$\f{1}{4}\hbar^2 \rho^2 { \bf j}^2$ and
observe that the following expression
\be
\hbar^2 \rho^2 \left[|\partial \chi_1|^2 +|\partial 
\chi_2|^2 - \f{{ \bf j}^2}{4}\right]
\la{comb}
\ee
is also gauge invariant.
Indeed if we introduce  the gauge invariant field
\be
\bfn = ({\bar \chi} \ , \ {\vec \sigma} \ \chi),
\ee
where $\chi =(\chi_1, {\chi_2^*})$ and $\vec \sigma$ are 
the Pauli matrices, then
$\bfn$ is a unit vector $|\bfn|=1$ and
we can write (\ref{comb}) as follows,
\be
\hbar^2 \rho^2 \left[|\partial \chi_1|^2 +
|\partial \chi_2|^2 - \f{{ \bf j}^2}{4}\right]=
\f{1}{4}\hbar^2 \rho^2  (\partial \bfn)^2
\ee

Consider now the remaining terms in (\ref{act}). 
For the magnetic field we get 
\comment{
We should remark that, in contrast to Abrikosov vortex (which 
is a topological defect which forms  in type-II 
superconductors in the presence 
of external manetic field) the  knotted solitons (which we discuss below)
are characterised by a nontrivial configuration of cupercurrent which 
is accompanied by induced magnetic field, thus in the effective action we 
express in the new varaibles the magnetic field
which appear in the presence of nontrivial supercurrent,
with {\it no} external magnetic field applied.}
\beg
{\bf B} = {\rm rot}{\bf A} = -\f{ c}{8  e}{\rm rot} 
\vec{\sf  C} + \f{\hbar c}{4 e} {\rm rot}{\bf j}
\la{hhh}
\eee
where ${\rm rot}\ \bf j$ can be written in terms of the
unit vector $\bfn$ as follows: 
\[
{\rm rot}{\bf j} = \f{1}{2}\bfn \cdot \partial_i
\bfn \times \partial_j\bfn 
\]
Combining these we then 
arrive at our main result:
The GLGP free energy density becomes
\beg
F&=& \frac{\hbar^2\rho^2}{4}(\partial \bfn)^2 + 
\hbar^2(\partial \rho)^2 +
\nonumber \\
&& \frac{\hbar^2 c^2}{512 \pi e^2}
\left(\f{1}{\hbar}[\partial_i {\sf C}_j -\partial_j 
{\sf C}_i] -\bfn \cdot \partial_i
\bfn \times \partial_j\bfn
\right)^2
\nonumber  \\ && 
+ \f{\rho^2}{16}
 \vec{\sf C}^2
+
{{V}}(\rho , n_3 ) 
\la{e3}
\eee
where we identify a version of the nonlinear O(3) $\sigma$-model 
introduced in \cite{fadde}
\be
F_0 \ = \  \frac{\hbar^2\rho^2}{4}(\partial \bfn)^2 + 
\frac{\hbar^2 c^2}{512 \pi e^2}
\left(  \bfn \cdot \partial_i
\bfn \times \partial_j\bfn
\right)^2
\la{fadmod}
\ee 
in interaction with a vector field $\vec{\sf C}$.
With this we have completed the 
mapping between the two-condensate 
GLGP model (\ref{act}) and (\ref{e3}), which is an extension
of the model (\ref{fadmod}) introduced in \cite{fadde}.
We emphasize that this involves an exact change
of variables between the GLGP model and (\ref{e3}).
This change of variables in particular eliminates 
the gauge field and as a consequence
the final result involves only the physically relevant field 
degrees of freedom that are present in the 
two-condensate system.

Note in particular the appearance of the 
mass for $ \vec{\sf C}^2$ which is a manifestation of the 
Meissner effect; the London magnetic field 
penetration length is  
$\lambda^2=\frac{c^2}{16\pi e^2}\left[ \f{|\Psi_1|^2}{m_1}+
\f{|\Psi_2|^2}{m_2}\right]^{-1}=\frac{c^2}{32\pi e^2\rho^2}$.
\comment{ The  term $\rho^2 \vec{\sf C}$ has a natural physical 
interpratation being  the kinetic energy density
of supercurrent $\bf J$ of Cooper pairs 
$W_{kin}=1/2(m_1/N_1+m_2/N_2)({\bf J}^2/e^2)$}
\comment{
Indeed we can see it 
explicitly  in the  London  limit.
That is we can take $\rm rot$ from both sides of (\ref{jjj})
and neglect contribution from the 
gradients of the order parameter ({\it London limit})
then taking into account 
the  Maxwell equation ${\bf J} = (c/4 \pi)  \ {\rm rot \ rot} {\bf A}$
the expression  (\ref{jjj}) yields the celebrated London equation:
\be
{\bf H} + \lambda^2{\rm rot \ rot} {\bf H} = 0
\label{london}
\ee
where as it can be checked by explicit calculations }
We also emphasize that the contribution $\bfn \cdot \partial_i
\bfn \times \partial_j\bfn$ to the magnetic field term 
in (\ref{e3}), (\ref{fadmod}) 
is a fundamentally important property of the 
two-condensate system which has no counterpart in a single
condensate system. Indeed, it is exactly
due to the presence of this
term that the two-condensate system acquires properties
which are qualitatively very different from those of a single-condensate
system: This term describes the magnetic field that
becomes induced in the system due to a nontrivial electromagnetic
interaction between the two condensates.

The potential term ${V}$ depends only on Cooper pair concentrations
and masses, 
and it is a function of the modulus $\rho$ and the  $n_3$ component  
of the vector $\bfn$ only.
In particular, we can write the mass term in (\ref{e3}) 
explicitely as follows:
\beg
{ V}&=& A + B n_3 + C n_3^2.
\la{e32}
\eee
Where
\beg
A&=&\rho^2 [  4c_1m_1^2 + 4c_2m_2^2 -b_1m_1 - b_2 m_2 ] \nonumber \\
B&=&  \rho^2 [ 8c_2m_2^2 - 8c_1m_1^2 -b_2m_2 + b_1 m_1] \nonumber \\
C&=& 4\rho^2 [c_1m_1^2 + c_2m_2^2]
\la{e32b}
\eee
This mass term determines the energetically preferred 
ground state value for $n_3$, which we denote 
by ${\tilde n_3}$. Explicitely,
\be
{ \tilde n_3} = \frac{\frac{N_1}{m_1} 
-\frac{ N_2}{m_2}}{\frac{  N_1}{m_1} + \frac{N_2}{m_2}}
\label{nstar}
\ee
Thus the ground state value of $\bfn$ is a circle specified   
by the condition $n_3={\tilde n}_3$
on the unit two-sphere.
This yields a 
uniform, unperturbed ground state for the condensates. 
Note that the ground state value 
only depends on the concentrations $N_\alpha$ and
the masses $m_\alpha$ of the Cooper pairs. 

However, here we are mostly interested 
in topological defects where the unit 
vector $\bfn$ locally deviates from this ground state value.  
This then corresponds to the formation of a local
inhomogeneity in the densities of the Cooper pairs. 
But in order to ensure a finite energy, at a certain 
distance from the inhomogeneity the 
vector $\bfn$ should return to the circle defined by (\ref{nstar}).

Indeed, the equivalence between the models
(\ref{act}) and (\ref{e3}) reveals a hidden topological 
structure in (\ref{act}) which has a number of important physical
consequences. In particular, it leads to the understanding 
of vortex types which are allowed by a two-condensate 
system: The model (\ref{fadmod})
is known to admit topological defects that have the shape  
of stable knotted solitons and are 
characterized by a nontrivial 
Hopf invariant \cite{fadde}-\cite{hie}, (see also comment \cite{vol}).
The simplest nontrivial solution to the equations of motion 
that follow from (\ref{fadmod}) was proposed in \cite{fadde}. In
terms of vector $\bfn$ it forms a toroidal vortex ring,  
twisted once around its core before joining the ends. 
The equivalence
 between the two-condensate
model (\ref{act}) and (\ref{e3}) then implies that the
GLGP model with two charged condensates not only possesses a 
hidden $O(3)$ symmetry, but should also allow for the formation of such 
knotted vortex solitons. The stability of these topological defects has 
been confirmed in extensive numerical simulations \cite{nature}-\cite{hie}
(for video animation of the  numerical simulations of 
the  knotted solitons  see the www-address indicated in Ref.  \cite{hie}).
Indeed, the knotted vortex solitons {\it e.g.} in a two-condensate 
superconductor should consist of finite-length 
stable closed vortices, which carry
a nontrivial helical configuration
of the magnetic field. These finite length closed vortices have
properties which are substantially different from those of a loop
which is formed by an ordinary Abrikosov vortex. In particular: the present
vortices are protected against shrinkage by the third term 
in (\ref{e3}) [second term in (\ref{fadmod})]. 
\comment{
A remarkable feature is that the model (\ref{e3}) has 
a well-defined limit (\ref{e4}) corresponding to extreme type-I
superconductors. This proves that the two-component type-I condensates 
allow knotted solitons, whereas it does not allow formation
Abrikosov vortices (see also comment \cite{type1}).}

Since the variable $n_3$ acquires the preferable value 
(\ref{nstar}) in the ground state,
the knotted solitons have the following form: At spatial infinity the 
unit vector $\bfn$ assumes a value on the circle which is 
defined by the condition $n_3 = {\tilde n_3 }$. We denote this value of 
$\bfn$ at spatial infinity by
\be
\bfn^\infty=(n_1^\infty, n_2^\infty, {\tilde n_3})      
\ \ \ \ \ \ \  (\bfn^{\infty}=\bfn( {\bf x } \rightarrow \infty)).
\ee
At the center of a knotted soliton the unit vector then
reaches a value which corresponds to the opposite 
point on the unit sphere. In the case of a nontrivial knotted soliton
the curve where the vector $\bfn$ coincides with this opposite point
value in general forms a closed loop or a knot. 
This is the {\it core} of the knotted soliton and we denote it by \cite{comme1}
\be
\bfn_0 = (-n_1^\infty, -n_2^\infty, -{\tilde n_3}) 
\ \ \ \ \ \ (\bfn {\rm \ value \ at \ core })
\ee
\comment{
The magnetic field 
apparently vanishes outside core but also
it vanishes in the knot soliton core. }
At the core the densities of the condensates
are characterized by the following non-vanishing values
$|\Psi_{1}(\bfn_0)|^2 =\frac{m_1}{m_2}N_2;  \  \
|\Psi_{2}(\bfn_0)|^2 =\frac{m_2}{m_1}N_1$.
In between the core where $\bfn=\bfn_0$ and the boundary 
where $\bfn \to \bfn^\infty$ the unit vector $\bfn$ 
in general rotates in a manner
which is determined by the ensuing 
Hopf invariant \cite{hopf}. Consequently a 
knotted solitons can be characterized by the number of 
rotations  which are performed by the component of $\bfn$, which
is perpendicular to the axis defined by the boundary 
conditions $\bfn^\infty$ and $\bfn_0$ when we move  
around the knotted soliton by covering it once in toroidal and 
once in poloidal directions over any surface which is located between 
the core $\bfn=\bfn_0$ and the boundary $\bfn=\bfn^\infty$.
In particular, in the region between the core and  the boundary the magnetic 
field acquires a contribution from $\bfn$ with a helical
geometry, ${\bf B}^h \propto \bfn \cdot \partial_i
\bfn \times \partial_j\bfn$.  

A schematic plot of a toroidal-shaped knot soliton is given on Fig. 1. 
\begin{figure}[htpb]
\vspace{0.1cm}
\epsfxsize=0.8\columnwidth
\centerline{
\epsffile{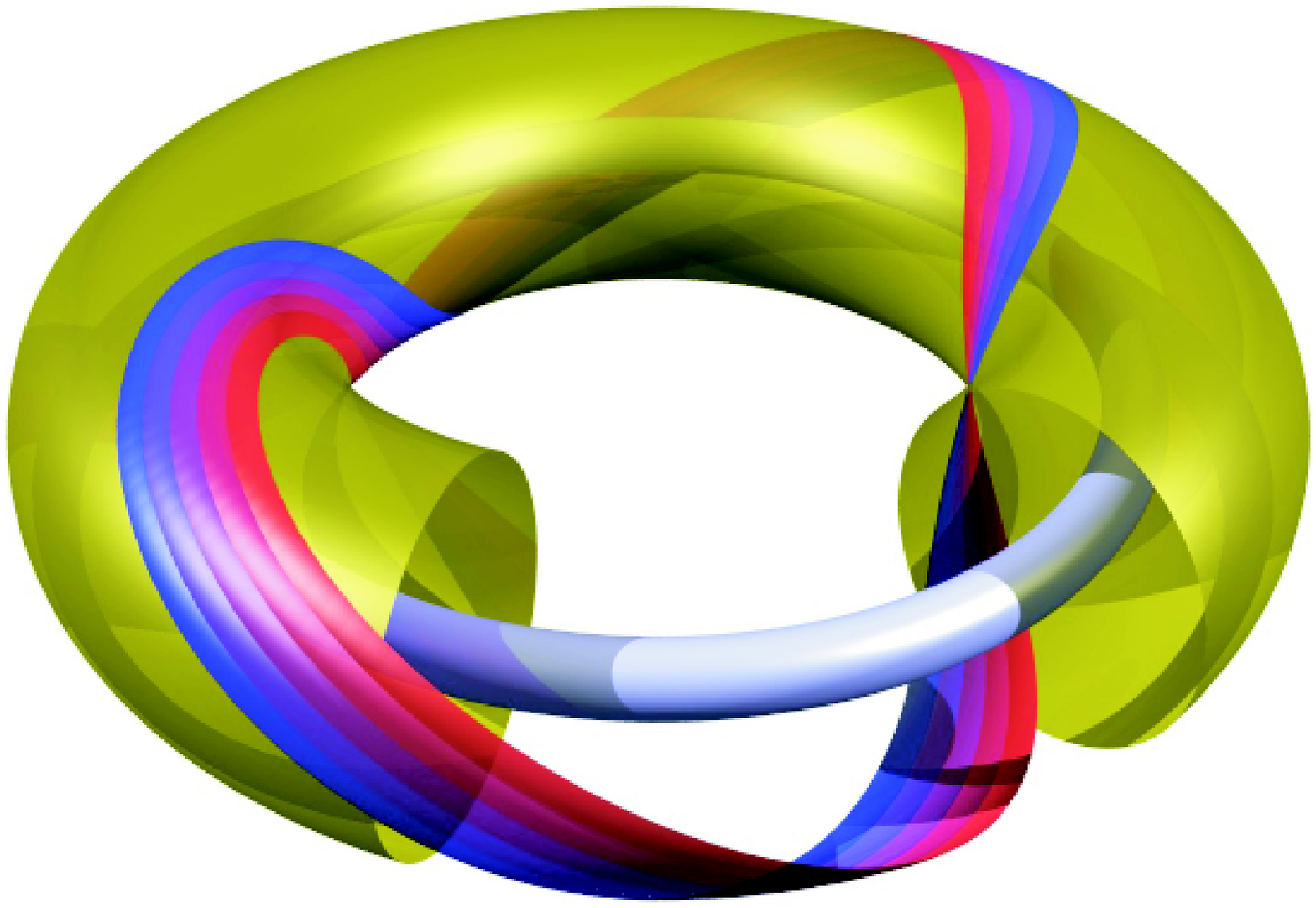}
}
\caption{ A schematic plot illustrating helical geometry 
of the vector $\bfn$  in a simplest toroidal knot soliton 
(for simplicity 
we neglected anisotropy effects connected with nonzero mass of the $n_3$
component of the vector $\bfn$).
The toroidal hollow surface  
is a surface situated between the knot core  $\bfn_0$
and the knot boundary $\bfn^\infty$. On this 
surface the vector $\bfn$ has
a constant projection to the axis defined by the boundary conditions
$\bfn^\infty$ and $\bfn$.
The spiral lines on this surface indicate the lines where  
the  vector $\bfn$ has some particular 
constant position on the unit sphere
 (the picture schematically  shows a toroidal knot
soliton which, in terms of the vector $\bfn$, was 
twisted twice around its core before joining its ends). If we proceed 
along an arbitrary path over this surface covering the knot soliton 
once in toroidal direction  and then  once in poloidal direction, the vector  
$\bfn$  
will perform $n$ and $ m $ rotations correspondingly,
according to the Hopf invariant which characterizes the given  soliton. 
In a general case the fluxtube can also be knotted (see also animations 
available at http address [9] ).
}
\label{exp}
\end{figure}

Finally we comment on the length scales that are present in our
system. For simplicity we assume that the two condensates have 
equal coherence lengths $\xi_1=\xi_2=\xi$. In this case one concludes
immediately that (\ref{e3}) has two different length scales. These are
$\lambda$ which is the mass of the field $\vec{ \sf C}$,  
and $\xi $ which is the mass of the component $n_3$ and relates to 
the coherence lengths of condensates. In the type-I limit which 
corresponds to $\lambda << \xi $, and in the type-II limit
where $\lambda >> \xi $ the characteristic sizes of knotted solitons 
are then determined by different factors: In the type-I limit 
the size is determined by an interplay of the terms 
$(\partial \bfn)^2$ and $(\bfn \cdot \partial_i \bfn \times 
\partial_j\bfn)^2$ which implies that the size is of order 
$\lambda$ \cite{type1}. In the type II limit the size of the soliton
is determined by the large mass of $n_3$ and thus it is of order 
of $\sqrt{\lambda \xi}$.

\comment{
An  interesting inherent feature
of the two condensate system with the prefereable 
value for $n_3$ is that besides
the magnetic field 
vanishes inside core and outside the knot, also in the limit (\ref{e4}) the
knot soliton 
features two closed  (or knotted and closed) spiral 
lines where magnetic field also vanishes. 
These lines  are winding around the core.
It can be verified by a straighforward  
calculation that these closed  spirals
corespond to the lines where 
the $\bfn$ vector reaches the  north and south poles of the unit spere. 
In the special case when  $N_1/m_1 = N_2/m_2$ these closed spiral lines
of zero magnetic field are situated at equal distance from the knot core but 
aparently never 
intersect each other.}
 
In conclusion, we have argued that charged two-condensate Bose systems
possess a number of interesting properties which are qualitatively
different from those of a single condensate system.
In particular they can
support stable knotted solitons as topological defects. Indeed,
we find it remarkable that a GLGP functional for a two-component 
charged superfluid becomes intimately related to the model introduced 
by one of the authors in  \cite{fadde}
which has been previously found to be relevant for strong 
interaction physics \cite{nature,cho}. This is a manifestation
of the universal character of the model, which appears to describe
a wide range of systems despite the differences 
in their physical origins. 
As a consequence the possibility of experimental investigation of 
the formation, properties and interactions of knotted solitons
in approproate superfluids ( e.g. $Mg B_2$,  $Nb$, $Ta$, $V$
, $Nb$-doped $Sr Ti O_3$)  along with numerical simulations,
could then shed light even on the properties of similar objects 
in nonabelian gauge theories of fundamental interactions. 

E.B. gratefully acknowledges numerous fruitful discussions 
with Prof. G. E. Volovik,
and Prof.  N.W. Ashcroft,  Dr. V. Cheianov, Prof. S. Girvin and Prof. 
A.J. Leggett for useful comments. 
L.F.   thanks L.P. Pitaevskii for  
discussions  and the Department for Theoretical 
Physics of the Uppsala University for hospitality. 
A.N. thanks Helsinki Institute of Physics 
for hospitality during
the completion of this work. We also thank M. L\"ubcke for plotting the figure.
E.B. has been supported by grant STINT IG2001-062 and the Swedish
Royal Academy of Science.
L.F. has been supported by grants RFFR 99-0100101,  INTAS 99-01705 and
CRDF grant RM1-2244.

\end{multicols}

\begin{thebibliography}{99}
\bibitem{rema}
In this Letter we consider two oppositely charged scalar fields. The 
result holds also for two electromagnetically interacting Bose fields 
of similar charge (charge inversion $e \rightarrow -e$ of the 
field $\Psi_i$ amounts to $|\Psi_i| e^{i \varphi_i}\rightarrow 
|\Psi_i| e^{-i \varphi_i}$). We also emphasis that the Coulomb 
interaction is assumed being screened  by free carries as it
is the case in superconductors,
so the fields are coupled only by means of magnetic field.
\bibitem{vol}
Here we should mention an earlier discussion of the 
topological defects characterised by a nontrivial Hopf invariant
in condensed matter physics.
The  topological defects   in ${}^3 He$
(where the $SO(3)$-symmetry is explicit)
characterized by a nontrivial Hopf 
invariant were first discussed  in  \cite{vo2} (for an excellent 
review and citations see also \cite{vol3}). The 
main distinction with knotted solitons considered in this paper 
is that the presence of condensate
coupling by magnetic field  
results in the apperance in the expression for the free energy 
of the terms $ \propto (\bfn \cdot \partial_i
\bfn \times \partial_j\bfn
)^2$. This in turn results in the 
fact that the knotted solitons are being protected against shrinkage 
by an energy barrier.
\bibitem{comme1}
We should note that in a superconductor where the condensates 
in different bands do not conserve independently, the 
effective potential in (\ref{act}) contains 
terms of the form $J[\Psi_1^*\Psi_2+\Psi_2^*\Psi_1]$ 
which describe the band coupling (see the  paper by Leggett in \cite{tg}).
Globally this term  breaks the remaining $O(2)$ symmetry
and the ground state of  $\bfn$ becomes reduced to  a point on $S^2$. 
In other words in that case we have that not only $n_3$ component
of $\bfn$ is massive but {\it all} components of it are massive.
This however does {\it not } significantly affect the knotted solitons 
discussed in our paper. The effect of the presence of the band coupling 
is that we should choose the appropriate boundary conditions so that 
 at spatial infinity from a knot soliton 
the vector $\bfn$ assumes a position corresponding to this
point. Then the effect of the presence of band coupling will
merely result in increase of the knot energy since the 
$O(2)$ boson is massive in this case,  but it does not affect 
stability of these defects similarly as it is not affected by the nonzero mass for  $n_3$.
\end{thebibliography}
\end{document}